\documentstyle[11pt,aaspp4]{article}
\def\mpc{\,h^{-1}{\rm Mpc}}
\def\kms{\,{\rm {km\, s^{-1}}}}
\begin{document}

\title{Cosmological Constraints on the Host Halos of GRBs}

\author{Wen Xu\footnote{Department of Physics and Astronomy, Arizona
State University, Tempe, AZ 85287} and
Li-Zhi Fang\footnote{Department of Physics, University of Arizona,
Tucson, AZ 85721}
}
 
\begin{abstract}
The recently observed bright optical transients(OT) of high redshift GRBs 
indicate that they are in a violent dynamical state. We think it is
reasonable to assume that the GRBs form in the environment of 
gravitationally collapsed halos of the cosmic matter field, and we investigate 
the basic parameters of the halos which are favored to host GRBs. If the 
harboring coefficient $f$ of GRBs per halo is weakly dependent on the 
mass of the halo, the redshift data of GRB OTs can yield significant 
constraints on the massive halos hosting GRBs. We show that, in the 
framework of popular cold dark matter (CDM) models, the GRB-favored 
environments are newly collapsed halos (i.e. their ages less than about 
$2 \times 10^9$ yr) with masses around $10^9$ $h^{-1}$ M$_{\odot}$. In 
this scenario, low redshift GRBs, if they exist, could not have the same cosmic 
origin as the high redshift ones. To fit with the observed rate of GRBs, 
we conclude that each GRB halo can host probably no more than one GRB 
event on average. This result implies that GRBs may be related to the merging 
of the halos.

\end{abstract}


\section{Introduction}

\bigskip

The cosmological origin of $\gamma$-ray bursts (GRBs) is favored by many recent 
observations (Meegan et al. 1992; Costa et al. 1997; van Paradijs et al. 1997;
Bond 1997; Fail et al. 1997; Metzger et al. 1997; Kulkarni et al. 1998; 
Djorgovski et al. 1998). The discovery of the bright optical transient (OT) 
associated with the GRB990123 (Odewahn et al. 1999a,b) at a redshift of 
z=1.6 (Kelson et al. 1999; Hjorth et al. 1999) added another cogent
piece of evidence 
for the cosmological distances of GRBs. These results also lead to an 
``energy crisis'' of about $3\times10^{54}$ ergs if $\gamma$-ray emission 
is isotropic (Kulkarni et al. 1999). It is difficult to make a model of the 
energy source by known mechanisms of radiation, despite the
many  models of the GRB 
energy source which have been proposed (see the references in Hartmann 1996; 
Piran 1998). Nevertheless, the energy of GRBs indicates
that they are related to 
some kind of violent process.  Related to this idea, we note the galaxy host of GRB990123 is 
highly irregular (Bloom et al. 1999, Fruchter et al. 
1999). This is evidence that the hosts of GRBs are in a violent dynamical state, 
such as gravitationally collapsed halos. Therefore, it would be reasonable to assume
that GRBs form in collapsed halos during cosmic gravitational 
clustering. A question is then: What are the basic properties of these halos?
Namely, in what environment (halos) are GRBs likely to form? We will try, in this 
{\it Letter}, to approach this question cosmologically.

The current situation of GRBs is in some ways
similar to that of QSOs in the early 1970s:
the mechanism of their radiation is uncertain, but they are believed to be 
cosmological objects. Past experience tells us that the cosmological aspects 
of QSOs
can be studied even in the absence of knowledge of the radiation mechanism.
For instance, the abundance, two-point correlation function, and 
redshift evolution
of QSOs has been successfully investigated in the framework of large scale
structure formation (Efstathiou \& Rees 1988, Nusser \& Silk 1993, Blanchard, 
Buchartet \& Klaff 1993, Bi \& Fang, 1997). In this approach QSOs were
assumed to 
be hosted by collapsed dark matter halos of the cosmic mass field, while the 
radiation and other hydrodynamic processes of QSOs were described by phenomenological 
parameters. Considering that the hydrodynamic processes are local, it is reasonable to 
assume that the hydrodynamic conditions may not be modulated by the density 
inhomogeneities on scale sizes larger than the size of the halo. In this case, the 
probability of a halo having a QSO should be the same for all halos (Fang \& Jing 
1998).
 
With the encouragement of the success of the cosmological approach to QSOs, 
we introduce a phenomenological parameter $f$ to describe the average 
number of GRBs hosted by a collapsed halo. Consequently, the GRB birthrate at 
redshift $z$ should be proportional to the birthrate of the considered
collapsed halos. Thus, by comparing the cosmologically predicted halos with the
observed redshifts of GRBs, one can find what halo environment 
is favored  by GRBs, i.e. one can find the
mass and age of the halos which are likely to host GRBs.

The current data on cosmological properties of GRBs are still poor. However,
popular models of structure formation, such as the cold dark matter
cosmogony, show that the redshift evolution of collapsed halos is a strong
function of the parameters of the halos. In such a case, even a few redshift data are
able to constrain the models. For instance, the data of three high
redshift galaxy clusters were found to effectively rule out models with
high mass density (Bahcall, Fan \& Cen 1997).  Therefore, one can expect that
the few redshift observations of GRBs would also be able to provide useful
constraints on the halos hosting GRBs. In the next two sections, using
the data on the redshifts of GRB OTs and the rate of GRBs, we 
find the basic parameters  of GRB-hosting halos in the 
framework of CDM cosmogony.

\section{Redshift-dependence of the birthrate of GRBs}

In the hierarchical structure formation models with Gaussian initial 
perturbation, the comoving number density of collapsed halos in the mass 
range $M$ to $M+dM$ can be calculated with the Press \& Schechter formalism 
(1974) as
\begin{equation}
n(M,z) dM= -\sqrt{\frac{2}{\pi}}\frac{\rho_0}{M^2} 
	\frac{\delta_c}{\sigma(M,z)} \exp 
	\left( \frac{-\delta_c^2}{2\sigma(M,z)^2} \right)
\frac{d\ln \sigma(M,z)}{d \ln M} dM,
\end{equation}
where $\delta_c\approx1.69$ almost independent of cosmologies.
$\sigma(M,z)$ is the linear theory $rms$ mass density fluctuation
in spheres of mass $M$ at redshift $z$ within a top-hat window of radius 
$R$, and is determined by the initial density spectrum $P(k)$ and 
normalization factor $\sigma_{8}=\Delta(8\mpc,0)$.  

We consider two popular representative CDM models - the standard CDM
(SCDM) and the flat low-density CDM (LCDM). The parameters of the models
(Hubble constant $h = H_0/(100 \kms$ Mpc$^{-1}$), mass density
$\Omega_0$, cosmological constant $\lambda_0$ and $\sigma_{8}$) are taken
to be $(0.5, 1, 0, 0.6)$ and $(0.75, 0.3, 0.7, 1)$ for the SCDM and LCDM,
respectively. These models provide a reasonable description of many
observational properties of large scale structures of the universe.

Six high redshift galaxies have been found to host GRBs so far.  Collapsed
halos must be massive enough to be favored by
the huge energy and violent activity of GRBs. Thus, we 
have to introduce the first parameter for the environment of GRBs: the mass 
scale $M_{GRB}$ of GRB hosting halos.
 From eq.(1), the number 
density of the collapsed halos formed by the era $z$ with mass greater than 
$M_{GRB}$ is given by
\begin{equation}
N(>M_{GRB}, z)=\int^{\infty}_{M_{GRB}} n(M,z)dM.
\end{equation}
Because $n(M,z)$ decreases rapidly with $M$, $N(>M_{GRB}, z)$ is actually
dominated by halos with mass $M \simeq M_{GRB}$.  

In the case of objects like galaxy clusters, there is a one-to-one
correspondence between an object and a halo with mass $M$ and radius $r$
(Xu, Fang \& Wu, 1998; Xu, Fang \& Deng 1999). Obviously, the
one-to-one identification is incorrect for the GRB as one massive
collapsed halo may host many GRBs in its history. Moreover, the number
of GRBs hosted by a halo of mass $M$ may be different for different halos, because
the formation of a GRB is not only determined by gravitational
parameters of the halos, but also by hydrodynamical processes. To reduce this
uncertainty, we consider that the hydrodynamical
processes are local. This is, the
relevant hydrodynamical conditions may not be modulated by the density
inhomogeneities on scales much larger than the size of the halos considered.
In this case, the average number of GRBs ($f$) hosted by a halo of mass $M$ does not
depend on the structures larger than the halos of eq.(2), and is also
redshift-independent. Thus, the total number of GRBs hosted by halos of
eq.(2) is $fN(>M_{GRB}, z)$. 

The birthrate of halos of mass $M>M_{GRB}$ at $z'$ is $d N(>M_{GRB},z')/dt$. 
A GRB may not form at the same time as the birth of a collapsed halo of mass 
$M$. In either stellar models or active galactic center models of GRBs, the 
formation
of a GRB on average has to be later than the birth of its host by $\tau$, which is
the time scale of the evolution from the birth of a collapsed halo to the
formation of GRBs in the halo. Obviously, $\tau$ is dependent on hydrodynamics, and
different for different halos. However, similar to $f$, $\tau$ can be
treated as a parameter if we are interested only in the mean rate of the GRB
formations over a mass scale much larger than the GRB halos themselves. Thus, a
halo of mass $M_{GRB}$, formed at redshift $z'$, will contribute on average $f$
GRB(s) at epoch $z$, and
\begin{equation}   
t(z)-t(z')=\tau.   
\end{equation}  
The GRB birthrate at redshift $z$ is then
\begin{equation}
\phi(z)=f\frac{dN(>M_{GRB},z')}{dt} \frac{1+z'}{1+z}.
\end{equation}
The last term comes from the time dilation between $z=z$ and $z=z'$.

Fig. 1 plots the birthrate $\phi(z)/f$ vs. $z$ in models SCDM
and LCDM for a given $M_{GRB} = 10^9$ $h^{-1}$ M$_{\odot}$.
For other values of $M_{GRB}$, the curves would basically be the same
except for the normalization. The parameter $\tau$ is
taken to be 0.1 to 5 Gyr.
The curves in Fig.1 have similar features: sharply rising at 
redshift $z_{r}$ and falling at $z_{f}$. Therefore, the nonzero
range from $z_{r}$
to $z_{f}$ is weakly affected by the redshift dependence
of $f$. The peak of the birth rate
shifts to smaller redshift for larger $\tau$. This is expected as the 
larger $\tau$ means the later formation of GRBs. 

The observational data of GRB redshifts have been obtained in
a couple of ways listed below with decreasing reliability:
(1) Derived from absorption line features in the spectrum of
optical transient. It includes the measurements of $z=$ 1.6
for GRB990123 (Kelson et al. 1999; Hjorth et al. 1999) and $z=$
0.835 for GRB970508 (Metzger et al. 1997). (2) Derived from host galaxy,
by assuming a physical association between GRB and its host galaxy.
It includes the redshift measurements of $z=0.835$ for GRB970508
(Bloom et al. 1998), $z=0.966$ for GRB980703 (Djorgovski et al. 1998),
$z=1.096$ for GRB980613 (Djorgovski et al. 1999), $z=3.418$ for
GRB971214 (Kulkarni et al. 1998), and $1.3\le z \le 2.5$ for GRB970228
(Fruchter et al. 1998). (3) Derived from a variety of objects other than
OTs or host galaxies. Examples are GRB980329 at $z\sim 5$ (Fruchter 1999),
GRB970828 at $z\sim 0.33$ (Yoshida et al. 1999) and GRB980425 at $z=0.0085$
(Galama et al. 1998).
                                               
Our major conclusion is based only on the two OT redshifts.  
The OT redshift of GRB990123 ($z=$ 1.6) is further considered as the mean
of redshift distribution, because it is not far away from the middle value
  of the five available host galaxy redshifts.
The redshift $z=$ 1.6 is shown in Fig 1. This redshift is outside 
the birth rate curves with parameters of $\tau >2 \times 10^9$ yr for 
SCDM, and $\tau >3 \times 10^9$ yr for LCDM, regardless of $M_{GRB}$. However, 
if we take $z=3.4$ of GRB971214 seriously, even $\tau >1 \times 10^9$ yr 
curves can be excluded.  These upper bounds of $\tau$  are
significantly shorter than the Hubble time $t_{Hubble}$ at redshift $z=1.6$.
($t_{Hubble}$=3.1, 5.5 Gyr, respectively for SCDM and LCDM). Therefore, 
GRBs are most likely to form in {\it newly} collapsed halos.

A generic feature of the birthrate of Fig. 1 is the deficiency of 
GRBs at low redshift. The redshift range $z_r - z_f$ of each birthrate
curve is generally rather narrow. 
 $z_r$ is larger than 0.1 for 
$\tau < 2 \times 10^9$ yr. Especially for the model LCDM, we have 
$z_r \gg 0.1$. Therefore, if the $z$=0.0085 supernova really corresponds to 
GRB980425 (Galama at al. 1998), the model LCDM will be in difficulty. 
Namely, GRB redshifts of $z = 1.6$ and 0.0085 can't be simultaneously contained within 
redshift range of $z_r - z_f$. The $z$=0.0085 
supernova-related GRB probably does not belong to the same type of
GRBs at high redshifts in our model. The redshift distribution
also supports the speculation that the soft $\gamma$-ray
 repeaters in our galaxy are caused by neutron stars, and are different 
from GRBs with cosmic distances (Piran 1998).

Fig. 2 plots the birthrate $\phi(z)/f$ vs. $z$ in models SCDM and LCDM for 
$\tau= 1$ and 2 Gyr, respectively for SCDM and LCDM.
The parameter $M_{GRB}$ is taken to be in the range $10^6$ -
$10^{10}$ $h^{-1}$ M$_{\odot}$. This figure also shows a sharply rising 
of the distribution at redshift $z_{r}$ and a sharply 
falling at $z_{f}$. The peaks of the birth rate
shift to smaller redshift for larger $M_{GRB}$, as the larger $M_{GRB}$
formed later.

In order to fit the range of $z_r \simeq 1$ (GRB980703) and $z_f
\simeq$ 3 (GRB971214), the mass $M_{GRB}$ of halos
must be larger than $10^7$ $h^{-1}$ M$_\odot$. Halos of
$M_{GRB} \leq 10^7$ $h^{-1}$ M$_\odot$ form too early, and it is difficult for
them to simultaneously
cover the $z \simeq$ 1 and $z \simeq$ 3 for any $\tau$. On the other
hand, M$_{GRB}> $ $10^{10}$ $h^{-1}$M$\odot$ halos form too late.
Therefore, the most likely mass range of GRB halos is 
$M_{GRB} \simeq 10^7 - 10^{10} h^{-1}$ M$\odot$. This mass range is at
the lower end of masses 
typical of today's galaxies. Therefore, these halos would be most
readily identified as the
progenitors of galaxies. This result is consistent
with the observed hosts, as the mass of GRB hosts is on the order of the 
mass of halos that can form galaxies by merging. It doesn't mean, however, that
the hosts of GRBs {\it must} be galaxies, because some massive halos may
not be involved in galaxy formation. For these halos GRBs are not associated
with galaxies.

\section{Number counts of GRBs}

We now consider the reasonableness of the environment parameters
with the observed rate of GRB. From the birthrate eq.(4), the all-sky,
observable number per year, of gamma ray burst events at the epoch of 
redshift $z$ is 
\begin{equation}
N_{GRB}(z)dz = \phi(z) (1+z) dV,
\end{equation}
where $dV(z)$ is the volume element in redshift space. The term $(1+z)$
comes from the time dilation between $z=0$ and $z=z$. Thus, the total
number of GRBs per year is
\begin{equation}
N_{GRB} \equiv \int N_{GRB}(z) dz = f \cdot N_{halo},
\end{equation}
where
\begin{equation}
N_{halo} \equiv 
\int \frac{dN(>M_{GRB},z')}{dt} (1+z') \frac{dV}{dz}dz.
\end{equation}
The observed GRBs per unit time is then
\begin{equation}
N_{obs}=s N_{GRB} \frac{\delta \Omega}{4 \pi}=
sf \frac{\delta \Omega}{4 \pi}N_{halo},
\end{equation}
where $s$ is a factor less than 1 to describe observational efficiency
of GRBs, $\delta \Omega/4 \pi$ is the correction of angular scale if the 
radiation is beamed.

The all-sky observed rate of GRBs is approximately one per day, i.e.
$N_{obs} \simeq 400$yr$^{-1}$. Thus, one can calculate the
observation-theory ratio
$N_{obs}/N_{halo}$, which is equal to $sf\delta \Omega/4 \pi$. Fig. 3 shows
how $N_{obs}/N_{halo}$ depends on $M_{GRB}$ and $\tau$. It is 
interesting to see that for all the considered parameter ranges 
of $M_{GRB}$ and $\tau$,
$N_{obs}/N_{halo}$ is less than, and even far less than one. 

Because $s \leq 1$ and $\delta \Omega/4 \pi \leq 1$, we have a
lower bound to $f$ given by
\begin{equation}
f >  \frac{N_{obs}}{N_{halo}} .
\end{equation} 
Since $M_{GRB}\geq 10^{7} h^{-1}$M$_{\odot}$ (\S 2) and 
$\tau \leq 2$ Gyr (SCDM) and 3 Gyr (LCDM), we have $f > 3\times 10^{-5}$
(LCDM) and $3 \times 10^{-6}$ (SCDM). 
 
There is some speculation that GRBs may 
constitute a unique homogeneous population of sources which has  few
selection effect(Piran 1998).  
If we further assume that $s$ is not much less than 1, then
$f \simeq N_{obs}/(N_{halo} \delta \Omega/4 \pi)$. Although our guesses
about the beaming of the $\gamma$-ray emission varing from
isotropic explosion $\delta \Omega/4 \pi = 1$ (e.g. Iwamoto 
et al. 1998) to highly beamed radiation  $\delta 
\Omega = 0.1 - 0.001$ (e.g. Hartmann 1996), $f$ is generally not 
larger than 1 for the majority of value of $M_{GRB}$ and $\tau$ shown in Fig. 3. 
Our model predicts one new born massive halo can contribute no more than one GRB.
A possible interpretation of this result is that the GRB phenomenon is related to 
halo merging. In the hierarchical structure formation scenario, massive 
halos formed from merging of less massive halos, i.e. each merging of
massive halos will produce a new halo. Thus, each GRB halo will undergo no 
more than once of merging of itself. Imaging of GRB990123 shows 
that the GRB is apparantly within an ongoing merger (Fruchter et al. 
1999).

Moreover, if most GRB-hosted halos merged into galaxies, each galaxy
with mass $10^{12} h^{-1}$M$_{\odot}$ formed, on average, from $10^5$ merging
of $10^7$ $h^{-1}$M$_{\odot}$ halos; or $10^2$ merging of $10^{10}$
$h^{-1}$M$_{\odot}$ halos. In other words, in the entire evolutionary
history of galaxy formation, a galaxy on average underwent about
$10^2$ - $10^5$ GRB-hosted halo mergings. On the other hand, an estimation
of the mean GRB number per galaxy in their evolutionary history found
$\sim 10^4/\delta \Omega$ (Hartmann 1996), which is consistent with our
cosmological estimation. 

\section{Conclusions}

The redshift data of GRB OTs yield significant constraints on the
massive halos hosting GRBs if the popular models of CDM cosmogony
are employed. We show that these data can be fitted by standard CDM models.
Our model predicts GRBs probably to be new members of the family of high redshift objects
which trace the cosmic large scale structures. We show that the 
halos forming GRBs are newly
collapsed  ($\tau < 2 \times 10^9$ yr) with masses scales around
$10^9$ $h^{-1}$ M$_{\odot}$. If the current record of GRBs is free
of observational
selection effects, the observed rate of GRBs requires that each GRB host
halo will host no more than one GRB event on average. This strongly implies a
possible relation between halo mergings and GRBs.

These conclusions are based on our assumption of a constant 
harboring coefficient $f$. More generally, $f$ may depend on the mass
$M$ of the collapsed halo. For instance, the capacity of harbhoring
GRBs can be made proportional to $M$ (i.e.
a linear relation:$f = constant \cdot M$). 
Since the abundance of halos $N(z)$ is a strong function of redshift around 
$z_r$ and  $z_f$, any weak $M$-dependent relation of $f$
will not change the conclusion about $z_r$ and $z_f$. In this 
case, the constraints to $M$ and $\tau$ are still valid. 

Using the parameters of the possible GRB halos, we predict the
redshift distribution of GRB OTs, as shown in Figs. 1 and 
2. This redshift distribution
is not sensitive to how $f$ depends on $M$. Therefore, our results will
be further tested in near future when enough redshift data are accumulated 
to give us a reliable redshift distribution of GRBs.

\acknowledgements

We would like to thank an anonymous referee for a report which improved 
the presentation of the paper. WX thanks H. Yan, T. Lu, R. Windhorst,
D. Burstein, R. Clawson for valuable discussions.

\newpage

\begin{figure}
\figcaption{The redshift evolution of the birth rate of GRBs.
The ratio $\phi(z)/f$ is shown so that the curves do not
depend on the choice of $f$. The 
solid and dashed lines are for LCDM and SCDM, respectively.
The parameter $\tau$ is taken to be 5, 3, 2, 1, 0.1 Gyr (LCDM)
and 5, 3, 1, 0.1 Gyr (SCDM) from left to right. The parameter
$M_{GRB}$ is fixed at $10^{9}\  h^{-1}$M$_{\odot}$. The redshift 
of GRB990123 is indicated as a landmark of $z$. The model 
curves with the best choice by the redshift data of GRBs are 
highlighted as thick lines.}
\label{fig1}
\end{figure}

\begin{figure}
\figcaption{ The redshift evolution of the birth rate of GRBs.
The solid and dashed lines are for LCDM and 
SCDM, respectively. The mass of GRB parent halos are taken to be
$10^6, 10^7, 10^8, 10^9, 10^{10} h^{-1}$M$_{\odot}$ from top to bottom 
curves, respectively. $\tau$ is taken to be $2\times10^9$ years for 
LCDM and $1\times10^9$ years  for SCDM.  The redshift of GRB990123 is
indicated as a landmark of $z$.}
\label{fig2}
\end{figure}

\begin{figure}
\figcaption{The ratio $N_{obs}/N_{halo}$ as a function of 
parameters $\tau$ and  $M_{GRB}$. The solid lines are for LCDM and 
dashed lines for SCDM. The parameter $\tau$ is equal to 
$5, 3, 2, 1, 0.1$ Gyr from top to bottom. The highlighted two lines 
are the same as the two in Fig.1.
}
\label{fig3}
\end{figure}

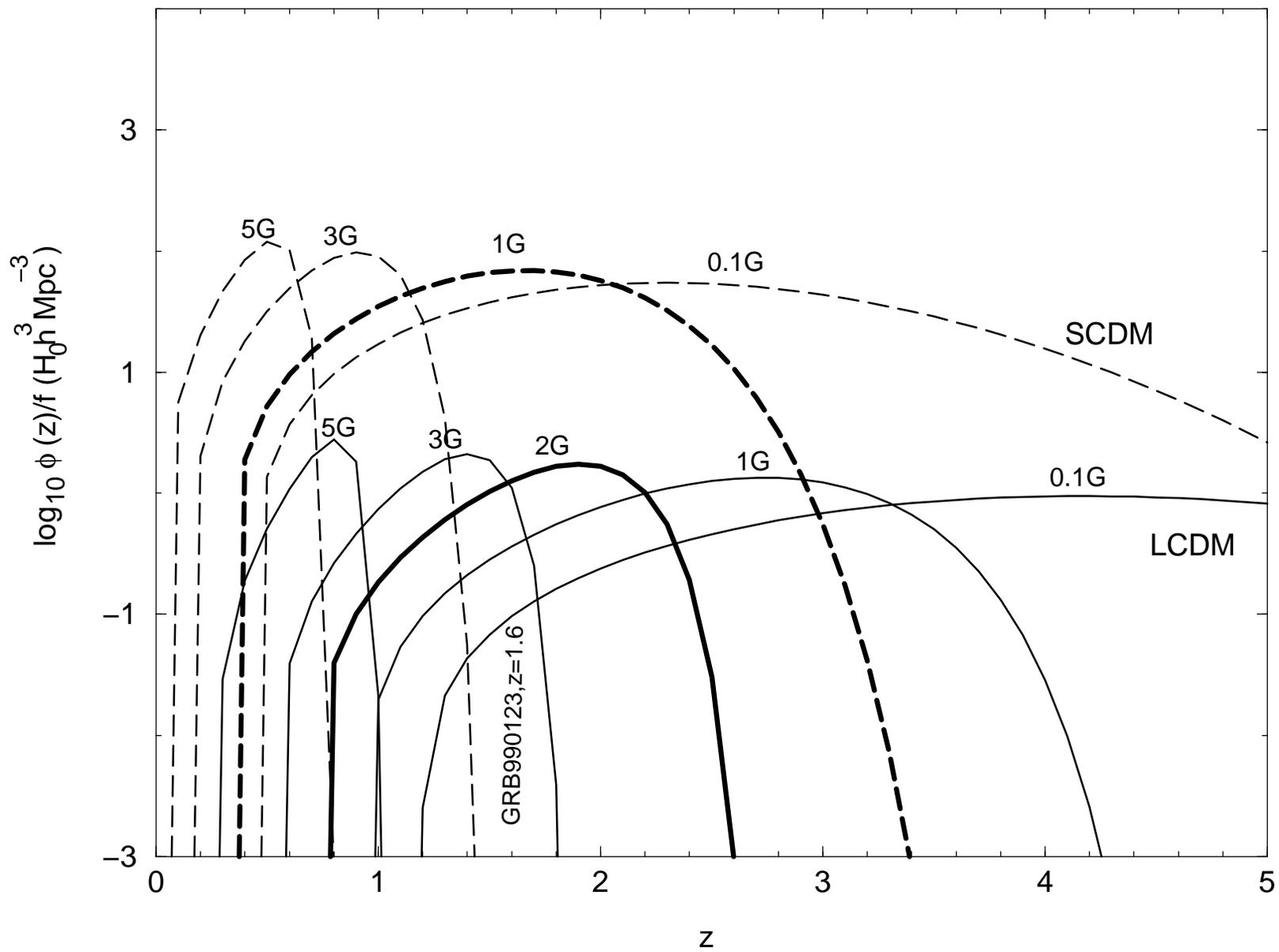
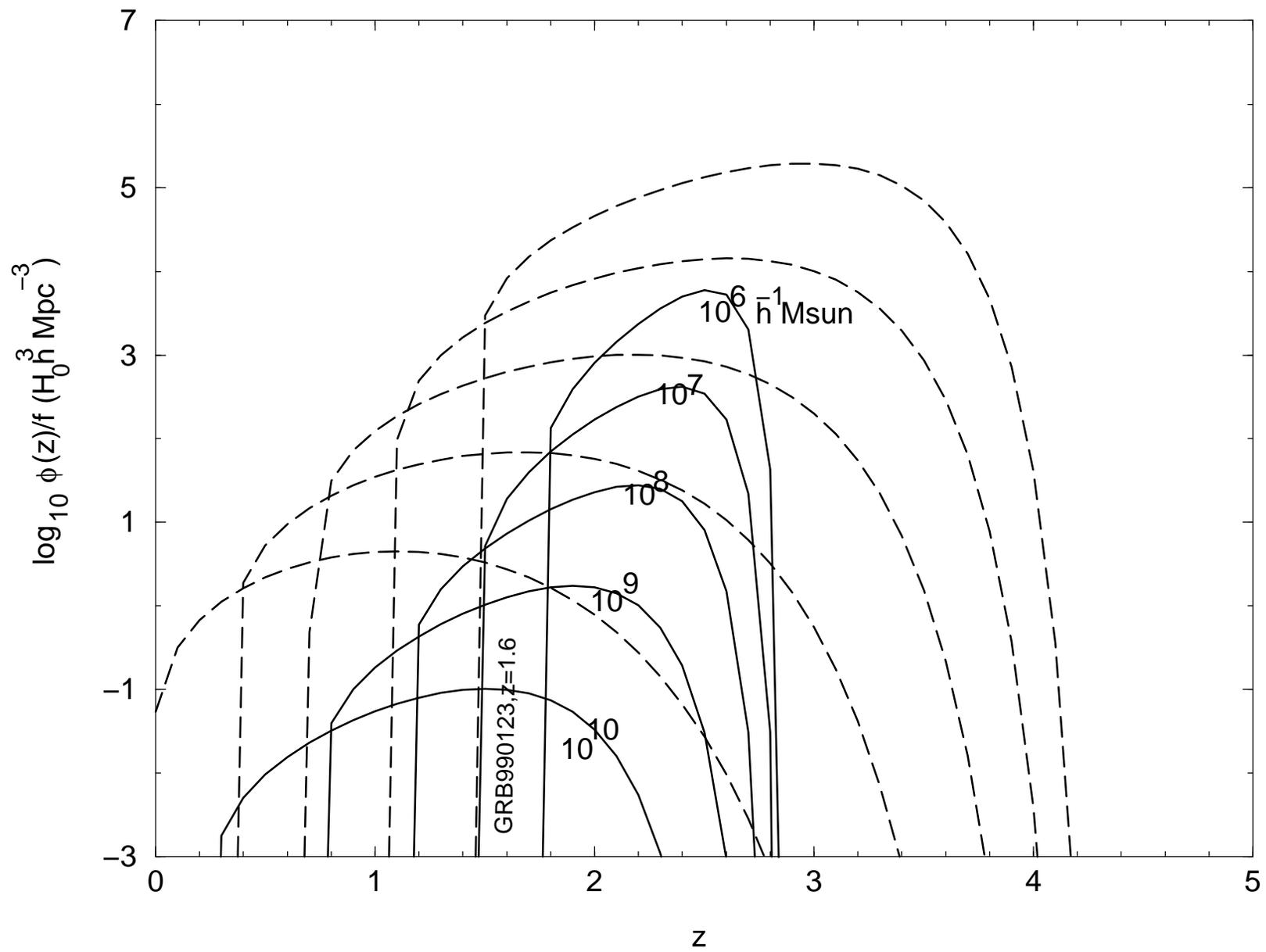
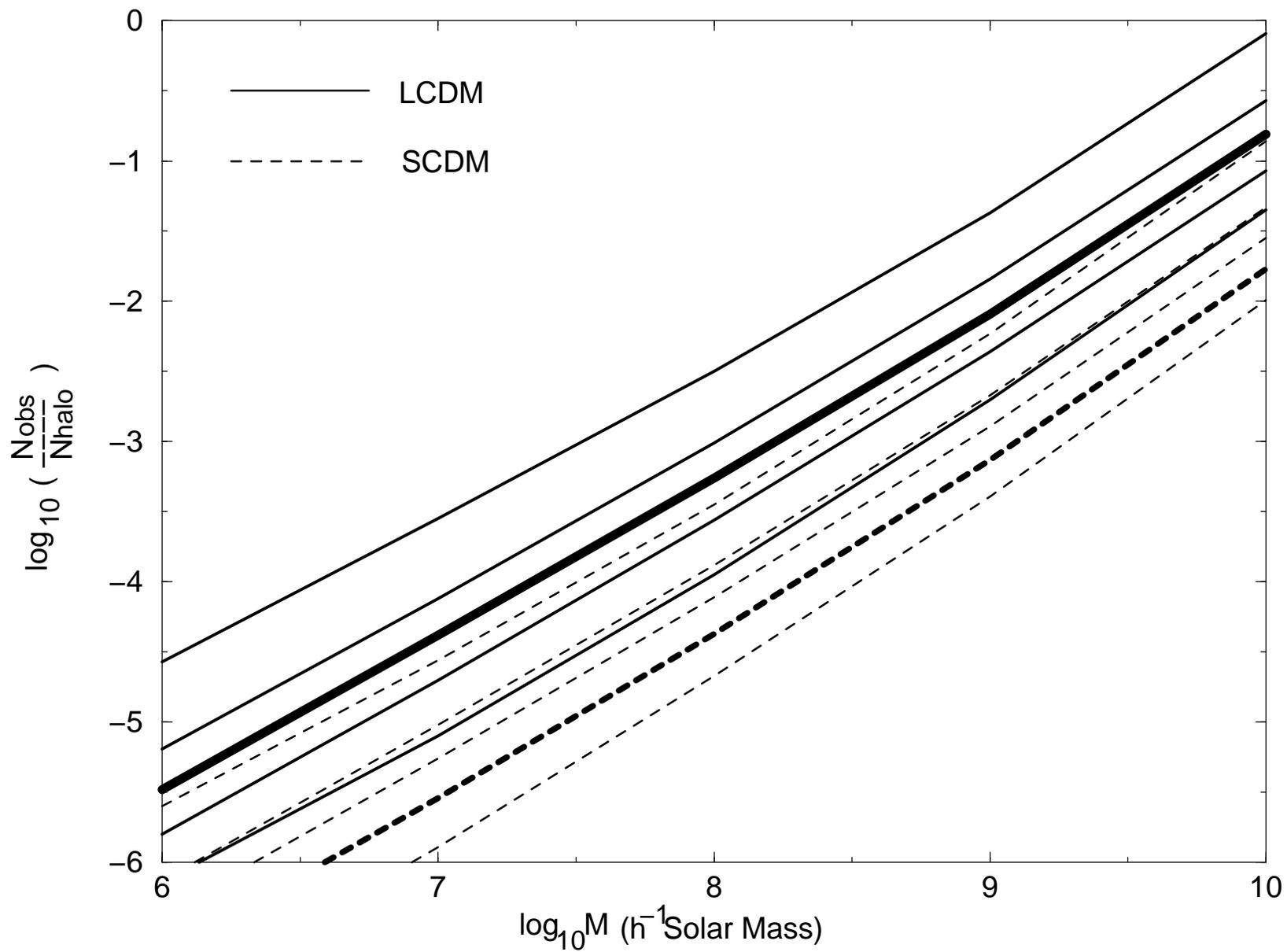
\end{document}